\documentclass[10pt,conference]{IEEEtran}
\IEEEoverridecommandlockouts

\usepackage{cite}
\usepackage{amsmath,amssymb,amsfonts}
\usepackage{algorithmic}
\usepackage{graphicx}
\usepackage{textcomp}
\usepackage{xcolor}
\usepackage{verbatim}
\usepackage{tikz}
\usetikzlibrary{arrows.meta,positioning,decorations.pathreplacing}
\usepackage{pbalance}
\usepackage{braket}

\def\BibTeX{{\rm B\kern-.05em{\sc i\kern-.025em b}\kern-.08em
    T\kern-.1667em\lower.7ex\hbox{E}\kern-.125emX}}

\makeatletter
\newcommand{\linebreakand}{%
  \end{@IEEEauthorhalign}
  \hfill\mbox{}\par
  \mbox{}\hfill\begin{@IEEEauthorhalign}
}
\makeatother

\begin{document}

\title{After Theft: From Revocation to Neutralization\\
in the Custody of Quantum Clones}

\author{%
\IEEEauthorblockN{Gabriele Gianini}
\IEEEauthorblockA{
\textit{University of Milano-Bicocca}\\
Milan, Italy \\
gabriele.gianini@unimib.it
}
\and
\IEEEauthorblockN{Jianyi Lin}
\IEEEauthorblockA{
\textit{Universit\`a Cattolica del Sacro Cuore}\\
Milan, Italy \\
jianyi.lin@unicatt.it}
\and
\IEEEauthorblockN{Corrado Mio}
\IEEEauthorblockA{
\textit{Khalifa University of Science and Technology}\\
Abu Dhabi, UAE \\
corrado.mio@ku.ac.ae}
\linebreakand
\IEEEauthorblockN{Stelvio Cimato}
\IEEEauthorblockA{
\textit{University of Milan}\\
Milan, Italy \\
stelvio.cimato@unimi.it}
\and
\IEEEauthorblockN{Ernesto Damiani}
\IEEEauthorblockA{
\textit{University of Milan}\\
Milan, Italy \\
ernesto.damiani@unimi.it}
}

\maketitle

\begin{abstract}
Encrypted quantum cloning enables the creation of multiple encrypted clones of an unknown quantum state while allowing only one effective decryption, with the decryption resource being intrinsically consumed in the process. In this paper, we argue that this property supports a distinctive custodial security primitive for post-compromise response. We consider a threat model in which an adversary steals an encrypted quantum clone but does not yet possess the corresponding decryption key. In such a scenario, a legitimate custodian may unseal a different trusted clone, thereby exhausting the sole available unsealing opportunity and rendering the stolen clone permanently useless. We argue that this mechanism is not adequately described as mere revocation. Rather, it realizes a stronger form of post-theft response, which we call neutralization. We formalize this distinction, locate it within a broader post-theft response space, and introduce a temporal threat model. We compare the mechanism with its closest classical analogue, showing that the classical case can reproduce the policy outcome only through external procedural composition. We finally interpret encrypted quantum cloning as a primitive for post-compromise quantum custody in distributed preservation settings, with prospective relevance for Cyber-Humanities-oriented preservation architectures.
\end{abstract}

\begin{IEEEkeywords}
Encrypted quantum cloning,
quantum custody,
post-compromise neutralization,
incident response,
controlled unsealing,
distributed preservation,
clone theft,
revocation.
\end{IEEEkeywords}

\section{Introduction}

Within Quantum Encrypted Cloning~\cite{yamaguchi2026encrypted,yamaguchi2026experimental,ceara2026cloningencryptedquantumstates}, multiple encrypted replicas of an unknown quantum state can be generated, but only one can be decrypted, consuming the decryption resource in the process. This differs from ordinary replication in classical distributed preservation systems~\cite{lockss_principles} and from ordinary access control~\cite{nist_access_control}: protected copies may remain distributed, yet meaningful unsealing remains available only once.
In this paper, we argue that this feature supports a distinctive \emph{post-compromise} (i.e., post-breach) custody primitive. 

The point matters whenever an encrypted quantum clone carries information of custodial relevance, either as the protected object of interest itself or as an access-bearing resource to it. We consider a threat model in which an \emph{adversary} steals an encrypted clone but does not yet possess the corresponding decryption key. A \emph{legitimate custodian} may then unseal a different trusted clone, thereby exhausting the sole available unsealing opportunity and rendering the stolen clone permanently useless. This possibility is granted by the non-local character of quantum information~\cite{he2026anomalous}.

This effect is stronger than ordinary \emph{revocation}. In the classical security literature, revocation is usually understood as the invalidation of future access authorizations, credentials, or tokens~\cite{nist_sp80053_revocation,nist_sp80063_token_revocation}. The mechanism studied here affects something else: not only future authorization, but the future effective decryptability of an already compromised dormant copy. We call this stronger effect \emph{neutralization}.

This work develops this distinction in a scoped way: we do not argue that quantum custody generally outperforms classical cryptography approaches to the problem; rather, we show that encrypted quantum cloning supports a custodial logic not natively reproduced by classical key destruction alone. 
\begin{comment}
More specifically, we locate revocation and neutralization within a broader post-theft response space, formalize the timing assumptions under which neutralization is feasible, compare the quantum mechanism with its closest classical emulation, and reinterpret the resulting logic as a primitive of post-compromise custody.
\end{comment}

Our contributions are the following. 
\begin{itemize}
\item 
We distinguish revocation from neutralization and argue that encrypted quantum cloning supports the latter as a native custodial response to clone compromise.
%\item
%We place this distinction within a minimal post-theft response taxonomy that includes detection, revocation, containment, and neutralization. 
\item
We introduce a temporal threat model centered on clone theft, compromise detection, key exposure, and legitimate emergency unsealing within a neutralization window. 
\item
We compare the mechanism with its closest classical analog, showing that the latter reproduces the policy outcome only through external procedural composition. 
\item
We reinterpret legitimate unsealing not only as an access event but also as a possible defensive act. 
\item
We discuss how this custodial primitive may be embedded in simple deployment architectures and why it is suggestive for distributed preservation 
in Cyber Humanities~\cite{adorni2025towards,Bellini2025Building,Bellini2025Cyber}.
\end{itemize}

As in standard quantum secret sharing models \cite{hillery1999quantum}, we treat share custody at
the level of access structures rather than physical storage technology:
authorized parties are assumed to control the shares, activation rights,
or resources required for unsealing, without committing the proposal to
any specific long-term quantum-storage implementation (see Appendix A).

The remainder of the paper is organized as follows. Section~II recalls the minimal background. Section~III introduces the distinction between revocation and \textit{neutralization} within a broader post-theft response space. Section~IV presents the temporal threat model. Section~V discusses the closest classical emulation and its limits. Section~VI interprets the resulting mechanism as a primitive of post-compromise quantum custody, discusses illustrative custody architectures, and outlines prospective Cyber-Humanities scenarios.

\section{Background and Motivating Scenario}

For the purposes of this work, protocol-level details of encrypted quantum cloning are not required: Appendix~\ref{app:single-use-unsealing} recalls the minimal formal structure on which the custodial argument relies. Here it is enough to adopt the following abstract picture. Let $|\psi\rangle$ denote an unknown quantum state, and let $\mathcal{C}$ denote an encrypted cloning procedure
\[
\mathcal{C}(|\psi\rangle) \longrightarrow \{c_1,\dots,c_n\}, K
\]
 producing a family of encrypted clones $c_1,\dots,c_n$ together with a decryption resource $K$. There exists an unsealing operation $U$ such that, for some clone $c_i$,
\(
U(c_i,K) \longrightarrow |\psi\rangle,
\)
while, after one effective unsealing event, the same resource no longer supports further recovery from the remaining clones.
Notice that, although we write the primitive in a qubit-like notation for simplicity, the custodial logic discussed here can be generalized to higher-dimensional quantum states.

The custodial significance of an encrypted quantum clone may arise in at least two ways. In a \emph{payload-bearing} interpretation, the clone protects the quantum state or the resource of direct interest. In an \emph{access-bearing} interpretation, by contrast, the clone protects a quantum resource whose primary role is to enable privileged access, authentication, validation, or unsealing with respect to some further protected object. In both cases, what matters is not only the confidentiality of the clone but the future utility that the compromised dormant clone may recover if the associated decryption resource is later exposed.

We consider a distributed-custody scenario in which multiple encrypted clones are preserved across distinct locations, one clone is stolen by an adversary, and the corresponding decryption resource has not yet been exposed. The stolen clone is unusable by itself, since encrypted possession is not equivalent to meaningful access. However, it still represents latent risk: if the decryption resource were later acquired by the adversary, the  stolen clone could in principle recover value.

The distinctive feature of the present setting is that a legitimate custodian may respond not only by revoking future access, but by unsealing a different trusted clone first, thus consuming the only effective unsealing opportunity. 
\section{From Revocation to Neutralization}

\begin{comment}
Before distinguishing revocation from neutralization, it is useful to locate both notions within a broader post-theft response space. At a high level, one may distinguish between \emph{detection}, which establishes that theft or exposure has occurred; \emph{revocation}, which removes future access capability; \emph{containment}, which limits further adversarial utility; and \emph{neutralization}, which extinguishes the future effective decryptability of the compromised dormant copy itself. 
\end{comment}
In this paper, our focus is on the distinction between revocation and neutralization, and on the claim that encrypted quantum cloning supports the latter as a native custodial response.
%
%At first sight, the post-theft response just described may resemble a form of revocation. However, that description is incomplete. What is affected is not only future authorization, but the future effective decryptability of an already compromised dormant copy. We therefore distinguish between \emph{revocation} and \emph{neutralization}.
%
We adopt the following working definitions.

\textit{Definition 1 (Revocation).}
\emph{Revocation} is the invalidation or removal of future access authorizations, credentials, tokens, or decryption capabilities associated with a protected copy~\cite{nist_sp80053_revocation}.

\textit{Definition 2 (Neutralization).}
\emph{Neutralization} is the elimination of the future effective decryptability of an already compromised clone through a legitimate unsealing event that exhausts the only remaining meaningful opening opportunity.

These definitions are asymmetric. Revocation is formulated at the level of authorization, neutralization at the level of the compromised object's future decryptability: the compromised copy is rendered sterile with respect to any future meaningful opening: neutralization targets the compromised copy itself, rather than only the surrounding authorization layer. 

\begin{comment}
Suppose that one clone, say $c_s$, has been stolen. If $c_i$ is a trusted clone with $i\neq s$, then the post-theft response may be expressed schematically as
\begin{equation*}
U(c_i,K)\Longrightarrow |\psi\rangle
\quad\text{and}\quad
\forall j\neq i,\; \neg \mathrm{DecryptableFuture}(c_j).
\end{equation*}
The result is not merely that future access is denied. 
%
The distinction matters also operationally. 
\end{comment}

In the setting considered here, neutralization offers several advantages over revocation alone: 
\begin{itemize}
\item
Neutralization reduces the residual post-theft risk. After revocation, a stolen  copy may still remain dangerous in a deferred sense: it persists as a dormant object whose value may revive if the corresponding decryption resource is later exposed. Neutralization is stronger because it aims to remove precisely that deferred value.
\item
Neutralization remains meaningful even when the stolen clone is offline and physically unrecoverable:\ it deprives it of future utility without physically recovering it. Revocation can invalidate permissions, but it does not alter the fact that the adversary already has an encrypted clone.
\end{itemize}
In addition, with neutralization, unsealing is no longer only just an access event: it also becomes a defensive act. Table~\ref{tab:revocation-neutralization} summarizes the distinctions.

\begin{table}[tb]
\caption{Revocation and neutralization in the post-theft setting.}
\label{tab:revocation-neutralization}
\centering
\begin{tabular}{p{0.24\linewidth}p{0.30\linewidth}p{0.30\linewidth}}
\hline
 & Revocation & Neutralization \\
\hline
Primary target & Authorization or decryption capability & Future decryptability of the clone \\
Object of concern & Access rights, credentials, services, keys & Dormant stolen copy \\
Effect & Future access is denied or disabled & Future meaningful opening is extinguished \\
Need to recover stolen copy & Not required, but  copy may retain latent value & Not required;  copy is sterilized by unsealing  \\
Role of opening & Access event & Access and defensive response \\
\hline
\end{tabular}
\end{table}

\section{Temporal Threat Model}

The advantage of neutralization over revocation is inherently temporal. This problem arises only if a legitimate custodian can still act after theft but before the corresponding decryption resource is exposed. 
%
\begin{comment}
Before formalizing the timing assumptions of the main scenario, it is useful to distinguish several compromise events that may affect the custody of the clones. The most important topic in this paper is \emph{clone theft}, in which an adversary acquires an encrypted clone while lacking the corresponding decryption resource. A second event is \emph{key exposure}, in which the decryption resource itself becomes available to the adversary. A third is \emph{insider compromise}, where an authorized custodian or institutional actor improperly enables access or exfiltration. A fourth is \emph{custodian failure}, where a legitimate response becomes impossible because trusted clones, custodial coordination, or unsealing capability are no longer available. Here we focus on the first case and model the second explicitly as the event that closes the neutralization window; the remaining cases are left as natural extensions of the threat model.
\end{comment}
%
We model the evolution of the system through four key times:
\begin{itemize}
\item $t_T$: time of clone theft;
\item $t_D$: time at which the legitimate custodian detects theft;
\item $t_U$: time at which a trusted clone is legitimately unsealed;
\item $t_K$: time at which the decryption resource $K$ is exposed to the adversary.
\end{itemize}
We assume throughout that the stolen clone remains encrypted at time $t_T$, that possession alone does not yield meaningful access, and that at least one trusted clone remains available for legitimate unsealing.
The central condition for successful neutralization is that legitimate unsealing must occur after detection but before key exposure:
\begin{equation}
t_T < t_D \leq t_U < t_K.
\label{eq:neutralization-window}
\end{equation}
When~\eqref{eq:neutralization-window} holds, the custodian can still act within the \emph{neutralization window}. If a trusted clone is unsealed during that interval, the stolen clone loses all future decryptability.

Beyond the ordering condition in~\eqref{eq:neutralization-window}, the practical viability of neutralization also depends on the operational width of the response window. Let
\(
W_N = t_K - t_D
\)
denote the interval between compromise detection and key exposure. If $\tau_U$ denotes the time required to authorize and execute legitimate unsealing on a trusted clone, then post-theft neutralization is operationally feasible only if
%\begin{equation}
\(
\tau_U \leq W_N.
\)
%\end{equation}
Here $\tau_U$ should be broadly understood as the response latency of legitimate unsealing, including detection confirmation, custodial coordination, authorization, and execution.

This makes clear that the advantage of neutralization depends not only on the existence of a response window but also on whether the custodian can act quickly enough within it:
the claimed advantage is conditional rather than absolute. If the key is exposed before detection or before response, then the stolen clone may still recover the adversarial value. Neutralization is therefore best understood as a form of \emph{time-sensitive post-compromise recoverability}.
To clarify the structure of the claim, it is useful to distinguish two representative cases.

\subsection{Case 1: Clone Theft Before Key Exposure}

In the favorable case, a clone is stolen at time $t_T$, the theft is detected at time $t_D$, and the decryption resource remains uncompromised until some later time $t_K$. If a trusted clone is unsealed at some $t_U$ satisfying~\eqref{eq:neutralization-window}, then the stolen clone is neutralized before the adversary ever gains access to the decryption resource. The stolen object is thus transformed from a latent future risk into a permanently sterile artifact.

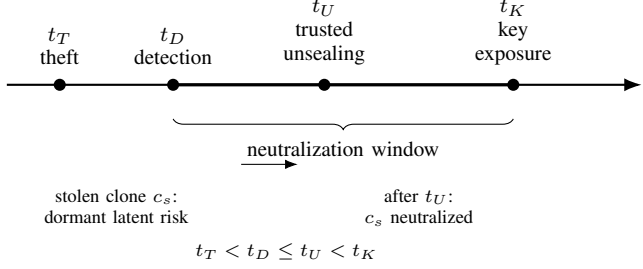
\begin{figure}[t]
\centering
\begin{tikzpicture}[
    >=Latex,
    font=\footnotesize,
    event/.style={circle, fill=black, inner sep=1.6pt},
    label/.style={align=center},
    window/.style={very thick},
    state/.style={align=center, font=\scriptsize}
]

\node[anchor=west, font=\footnotesize\bfseries] at (0,1.35)
{Successful neutralization};

\draw[->, thick] (0,0) -- (8.4,0);

\coordinate (tT) at (0.7,0);
\coordinate (tD) at (2.2,0);
\coordinate (tU) at (4.2,0);
\coordinate (tK) at (6.7,0);

\draw[window] (tD) -- (tK);
\draw[decorate,decoration={brace,amplitude=4pt,mirror}]
    (2.2,-0.45) -- (6.7,-0.45)
    node[midway,below=5pt,align=center] {neutralization window};

\node[event] at (tT) {};
\node[event] at (tD) {};
\node[event] at (tU) {};
\node[event] at (tK) {};

\node[label, above=4pt] at (tT) {$t_T$\\theft};
\node[label, above=4pt] at (tD) {$t_D$\\detection};
\node[label, above=4pt] at (tU) {$t_U$\\trusted\\unsealing};
\node[label, above=4pt] at (tK) {$t_K$\\key\\exposure};

\node[state, below=1.25cm] at (1.45,0)
    {stolen clone $c_s$:\\dormant latent risk};
\node[state, below=1.25cm] at (5.45,0)
    {after $t_U$:\\$c_s$ neutralized};

\draw[->, thin] (3.1,-1.05) -- (3.85,-1.05);

\node[align=center, font=\scriptsize] at (3.7,-2.20)
    {$t_T < t_D \leq t_U < t_K$};

\end{tikzpicture}
\caption{Temporal structure of post-theft neutralization. If a trusted clone is unsealed within the neutralization window, the stolen clone loses future effective decryptability.}
\label{fig:neutralization-window}
\end{figure}

\subsection{Case 2: Delayed Detection}

The second case arises when theft is detected too late. Suppose that 
%\begin{equation}
\(
t_T < t_K \leq t_D.
\)
%\end{equation}
Here, the adversary acquires the decryption key before the custodian has recognized the theft and responded. The neutralization window is closed by the time the response begins. 

This case shows that the proposed primitive does not eliminate the need for monitoring, audit, or timely detection: its usefulness depends crucially on how quickly compromise is recognized and acted upon.

\section{Classical Emulation and Its Limits}

The central comparison in this paper is not between quantum and classical security in general but between two ways of achieving post-theft containment. In the quantum-clone setting, legitimate unsealing may neutralize a stolen dormant clone because the decryption opportunity is intrinsically single-use. In a classical setting, the closest analog is procedurally composite: one decrypts a legitimate copy and then destroys, revokes, or disables the corresponding decryption resource.

\subsection{The Closest Classical Analogue}

Consider a classical protected object $o$ together with multiple encrypted copies
\(
E_K(o)_1,\dots,E_K(o)_n,
\)
where $E_K$ denotes encryption under key $K$. Suppose that one encrypted copy, say $E_K(o)_s$, is stolen while the key remains under legitimate control. A natural classical response is then:
\begin{enumerate}
\item decrypt a trusted copy $E_K(o)_i$ with $i\neq s$ (this assumes at least one trusted copy is still available);
\item destroy, revoke, or permanently disable the key $K$.
\end{enumerate}
If this sequence succeeds before the adversary acquires $K$, then the stolen ciphertext remains unusable in practice.

This prevents an overstated claim. The present paper does \emph{not} argue that only quantum systems can support a policy in which one retained copy is opened while a stolen encrypted copy becomes practically useless. A classical system can emulate that outcome under favorable assumptions.

\subsection{Policy Emulation Is Not Semantic Equivalence}

However, reproducing the same operational outcome does not amount to realizing the same custodial primitive. In the classical case, containment is externally composed out of at least two distinct actions:
\begin{equation}
\text{decrypt trusted copy} + \text{destroy or revoke }K.
\end{equation}
The first yields access; the second is an additional administrative or procedural step intended to prevent future access elsewhere.
In the quantum-clone setting, by contrast, containment is internal to the semantics of unsealing itself.
The same act that yields legitimate recovery also exhausts the future effective decryptability of the remaining dormant clones. The classical case therefore reproduces the \emph{effect} of neutralization, but not its \emph{native semantics}.

\begin{comment}
\begin{table}[tb]
\caption{Classical emulation and quantum neutralization compared.}
\label{tab:classical-quantum-comparison}
\centering
\begin{tabular}{p{0.26\linewidth}p{0.29\linewidth}p{0.29\linewidth}}
\hline
 & Classical case & Quantum-clone case \\
\hline
Target of defense & Authorization/key resource & Compromised dormant clone \\
Containment mechanism & Decrypt one copy, then revoke/destroy key & Legitimate unsealing intrinsically exhausts future opening \\
Residual post-theft risk & Latent value persists if key later leaks & Future effective decryptability is extinguished \\
Meaning of opening & Access plus separate administrative step & Access and containment in one act \\
\hline
\end{tabular}
\end{table}
%
%
Table~\ref{tab:classical-quantum-comparison} shows that the relevant contrast is not between classical insecurity and quantum security, but between two different post-theft custodial logics: externally coordinated containment and intrinsically single-use neutralization.
For this reason, encrypted quantum cloning is more naturally interpreted here as a custody primitive than merely as an encryption method.
\end{comment}

\section{Post-Compromise Quantum Custody in Distributed Preservation Settings}

In distributed preservation settings, redundancy is normally desirable:
multiple protected copies reduce the risk that a single institutional,
technical, or physical failure destroys access to the preserved object.
However, at the same time, it increases the adversary attack surface. Encrypted
quantum cloning is relevant to this tension because it \textit{separates
preservation redundancy} from \textit{unrestricted future unsealing}. With this respect it displays three relevant features:
\begin{itemize}
\item
\emph{dormant redundancy}: several protected copies may exist without
creating several independent future openings. 
\item
\emph{defensive
unsealing}: opening a trusted clone can also function as incident
response. 
\item
Third, \emph{post-compromise sterility}: after legitimate
unsealing, stolen dormant copies become
non-informative rather than merely administratively revoked.
\end{itemize}
%
\begin{comment}
Read as an incident-response primitive, encrypted quantum cloning suggests the minimal post-theft response sequence shown in Fig.~\ref{fig:post-theft-playbook}.

\begin{figure}[t]
\centering
\fbox{%
\parbox{0.95\linewidth}{%
\textbf{Minimal Post-Theft Response Pattern}

\begin{enumerate}
\small
    \item \textbf{Establish} that clone theft or suspected compromise has occurred.
    \item \textbf{Assess} whether the decryption resource remains unexposed and whether a trusted clone is still available for response.
    \item \textbf{Trigger} legitimate emergency unsealing on the trusted clone.
    \item \textbf{Thereby neutralize} the compromised dormant clone by exhausting the sole effective unsealing opportunity.
\end{enumerate}
}}
\caption{Minimal post-theft response sequence enabled by encrypted cloning.}
\label{fig:post-theft-playbook}
\end{figure}
\end{comment}
Taken together, these properties motivate the claim that encrypted quantum cloning supports a custody primitive rather than merely a confidentiality mechanism. The term \emph{custody}  
is used here in continuity with established archival theory, rather than as a merely technical security label \cite{saa_custody}.
%\cite{saa_custody,saa_custodial_history,saa_provenance,saa_chain_of_custody}.

The Cyber-Humanities relevance of this distinction becomes clearer in settings where preservation and access must remain sharply differentiated. The most relevant scenario is that of \emph{sealed distributed archives}, in which multiple institutions preserve dormant protected replicas while unrestricted future opening must remain preventable even after the theft of a copy.

\begin{figure*}[t]
\centering
\begin{tikzpicture}[
    font=\footnotesize,
    >=Latex,
    panel/.style={
        draw,
        rounded corners=4pt,
        inner sep=8pt,
        fill=gray!4
    },
    title/.style={
        font=\bfseries,
        anchor=west
    },
    custodian/.style={
        draw,
        rounded corners=3pt,
        minimum width=1.95cm,
        minimum height=0.75cm,
        align=center,
        fill=white
    },
    authority/.style={
        draw,
        rounded corners=3pt,
        minimum width=2.35cm,
        minimum height=1.0cm,
        align=center,
        fill=gray!12
    },
    governance/.style={
        draw,
        rounded corners=3pt,
        minimum width=1.65cm,
        minimum height=0.68cm,
        align=center,
        fill=white
    },
    adversary/.style={
        draw,
        dashed,
        rounded corners=3pt,
        minimum width=2.0cm,
        minimum height=0.78cm,
        align=center,
        fill=white
    },
    solidarrow/.style={->, thick},
    dashedarrow/.style={->, thick, dashed},
    label/.style={font=\scriptsize, align=center, fill=gray!4, inner sep=1pt}
]

% =========================================================
% LEFT PANEL: CENTRALIZED AUTHORITY
% =========================================================
\node[panel, minimum width=7.65cm, minimum height=5.5cm, anchor=north west] 
    (leftpanel) at (-2.0,2.9) {};
\node[title] at (-1.0,2.52) {(a) Centralized unsealing authority};

% Custodians
\node[custodian] (c1) at (0.0,1.55) {Custodian A\\encrypted clone $c_1$};
\node[custodian] (c2) at (0.0,0.35) {Custodian B\\encrypted clone $c_2$};
\node[custodian] (c3) at (0.0,-0.85) {Custodian C\\encrypted clone $c_3$};

% Authority and adversary
\node[authority] (auth) at (3.65,0.8) {Central\\unsealing authority\\key resource $K$};
\node[adversary] (adv) at (3.65,-1.05) {Adversary\\stolen clone $c_s$};

% Arrows
\draw[solidarrow] (auth.west) -- (c1.east)
    node[midway, above, sloped, label] {authorized\\unsealing};
\draw[solidarrow] (auth.west) -- (c2.east);
\draw[solidarrow] (auth.west) -- (c3.east);

\draw[dashedarrow] (c3.east) -- (adv.west) node[midway, below, sloped, label] {compromise};

% Caption note inside panel
\node[align=center, font=\scriptsize, text width=5.0cm] at (0.5,-2.1)
    {Distributed encrypted clones; one authority controls emergency unsealing.};

% =========================================================
% RIGHT PANEL: FEDERATED
% =========================================================
\node[panel, minimum width=9.05cm, minimum height=5.5cm, anchor=north west] 
    (rightpanel) at (6.15,2.9) {};
\node[title] at (7.95,2.52) {(b) Federated emergency unsealing};

% Custodians
\node[custodian] (fc1) at (7.85,1.55) {Custodian A\\encrypted clone $c_1$};
\node[custodian] (fc2) at (7.85,0.35) {Custodian B\\encrypted clone $c_2$};
\node[custodian] (fc3) at (7.85,-0.85) {Custodian C\\encrypted clone $c_3$};

% Governance nodes
\node[governance] (g1) at (10.45,1.7) {Archive A\\share $K_A$};
\node[governance] (g2) at (10.45,0.35) {Archive B\\share $K_B$};
\node[governance] (g3) at (10.45,-0.6) {Archive C\\share $K_C$};

% Federated unsealing and adversary
\node[authority] (fed) at (13.65,0.35) {Federated\\emergency\\unsealing};
\node[adversary] (fadv) at (13.65,-1.75) {Adversary\\stolen clone $c_s$};

% Arrows from custodians to governance shares
\draw[solidarrow] (fc1.east) -- (g1.west);
\draw[solidarrow] (fc2.east) -- (g2.west);
\draw[solidarrow] (fc3.east) -- (g3.west);

% Arrows from governance to federation
\draw[solidarrow] (g1.east) -- (fed.west)
    node[midway, above, sloped, label] {joint\\authorization};
\draw[solidarrow] (g2.east) -- (fed.west);
\draw[solidarrow] (g3.east) -- (fed.west);

% Stolen path
\draw[dashedarrow] (fc3.east) to[out=0,in=180] (fadv.west)
    node[midway, below, sloped, label] {};

% Caption note inside panel
\node[align=center, font=\scriptsize, text width=5.0cm] at (9.0,-2.05)
    {Dormant clones remain distributed; emergency unsealing requires federated authorization.};

\end{tikzpicture}
\caption{Two possible custody architectures for post-compromise quantum
neutralization. In the centralized model, a single authority controls the
key resource used for emergency unsealing. In the federated model, shares
of the unsealing resource are distributed among multiple custodial authorities, whose authorized collaboration is required to
activate emergency unsealing according to the chosen access structure.
Dashed arrows indicate compromise of  encrypted clone.}
\label{fig:custody-architectures}
\end{figure*}
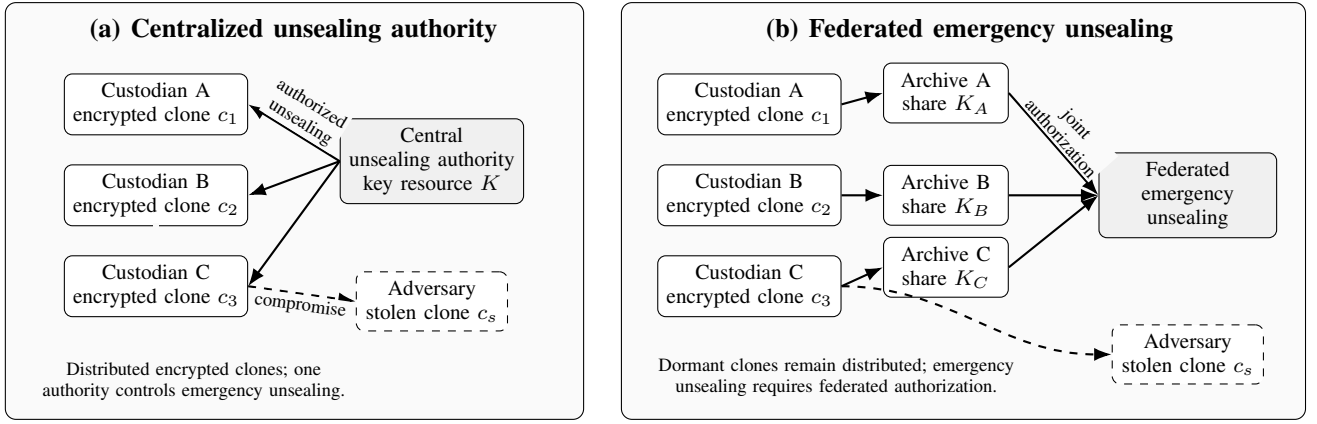

\subsection{Illustrative Custody Architectures}

The custodial logic developed here can be embedded in different deployment architectures. Two minimal models are especially useful for interpretation.
See Figure~\ref{fig:custody-architectures}.

\textit{Centralized unsealing authority.}
In the simplest model, encrypted quantum clones are distributed across multiple custodial locations, while the decryption resource and the authority to trigger legitimate unsealing remain centralized. This architecture makes the post-theft logic especially transparent: dormant redundancy is distributed, but the capacity to neutralize a stolen clone is concentrated in a single trusted domain. Its main advantage is conceptual simplicity; its main limitation is that the response surface depends strongly on the resilience and trustworthiness of the central authority.

\textit{Federated emergency unsealing.}
In a second model, encrypted clones remain distributed, but the authority to initiate legitimate emergency unsealing is itself governed across multiple custodians or institutions. Here, post-theft neutralization becomes not only a technical capability but also a matter of custodial coordination. This architecture is especially suggestive for inter-institutional preservation settings, but it also makes response latency more complex, since the feasibility of neutralization depends on organizational approval and coordination in addition to technical timing.
In this sense, the response latency $\tau_U$ introduced in Section~IV is not purely technical: it may reflect institutional coordination  and governance structure as much as protocol execution time.

\subsection{Prospective Cyber-Humanities Scenarios}

The relevance of the proposed distinction becomes clearer in Cyber-Humanities settings where preservation, access, and provenance must remain sharply differentiated.

A first prospective scenario is that of \emph{sealed distributed archives}, in which multiple institutions preserve dormant protected replicas while unrestricted future opening must remain preventable even after the theft of one copy. In such a setting, post-theft neutralization is attractive because it allows compromise response without requiring physical recovery of the stolen replica.

A second prospective scenario is that of \emph{federated inter-institutional custody}. Here, different cultural or archival custodians share preservation responsibility across institutions, while theft or exfiltration may occur locally. In this setting, neutralization is relevant because it supports a form of post-compromise containment compatible with distributed trust and distributed preservation.

%A third prospective scenario is that of \emph{provenance-sensitive or rare-access holdings}, in which opening a protected object is not merely a technical operation of access, but a rare, accountable, and curatorially meaningful event. Under such conditions, legitimate unsealing may matter not only because it grants access but also because it changes the future status of other dormant replicas through neutralization.

%These scenarios are intentionally prospective rather than operational. Their role is not to claim immediate deployment, but to indicate why the custodial logic developed in this paper may be especially suggestive for Cyber-Humanities-oriented preservation architectures.

Future work may also explore whether Cyber-Humanities could eventually
include \emph{native quantum cultural artifacts}: cultural objects whose
identity is not merely protected by quantum means, but partly constituted
by a quantum state or by an access-bearing quantum resource. 
This
possibility is consistent with emerging discussions of quantum humanities,
quantum aesthetics, and quantum-computing-inspired art
\cite{botticher2022quantumhumanities,liu2026quantumaesthetics}.
In such cases, the post-compromise custody logic discussed here may
acquire a more direct domain-specific significance, since the protected
resource would no longer be only an external security layer but part of
the conditions under which the object remains authentic, unique, or
performatively realizable.

\subsection{When Neutralization May Be Limited or Undesirable}

The proposed primitive also comes with important limitations. Its
usefulness depends on the timing assumptions formalized in
Section~IV: if key exposure occurs before compromise detection or before
legitimate response, neutralization may no longer be possible. It is also
not universally preferable to revocation or conventional access control.
When the protected asset is classical, when mature key-management
mechanisms are sufficient, or when repeated legitimate recovery from
multiple replicas is a primary requirement, conventional methods will
remain simpler and often more appropriate. Similarly, if response latency
dominates the threat model, the practical value of neutralization may be
limited.

These limitations do not weaken the conceptual distinction between
revocation and neutralization. Rather, they clarify the conditions under
which neutralization becomes meaningful: where post-theft containment
matters, where single-use unsealing is acceptable, where extinguishing
the future utility of a compromised dormant copy is preferable to
preserving broad recoverability, and where timely response remains
operationally feasible. The contribution of the present analysis is
therefore not universal replacement, but conceptual extension: it
identifies a custodial response that becomes available when encrypted
quantum cloning is taken seriously as a post-compromise primitive.

\section{Conclusions}

This paper has argued that encrypted quantum cloning supports a post-theft custodial response stronger than revocation alone. In the setting considered here, legitimate unsealing of a trusted clone may function not merely as access, but as \emph{neutralization}: it can extinguish the future effective decryptability of an already compromised dormant copy.

The distinction between revocation and neutralization is conceptually and operationally significant. Classical systems may emulate a similar policy outcome through decryption followed by key destruction, but they do not internalize post-theft containment into the semantics of unsealing itself.

This difference is especially suggestive for distributed preservation settings in which dormant redundancy, controlled opening, and post-compromise containment must coexist. Under those conditions, encrypted quantum cloning is best understood not merely as a storage or protection mechanism, but as a primitive for post-compromise quantum custody.

%More broadly, the custody-oriented interpretation proposed here is not inherently limited to qubit payloads, and may become even more suggestive in higher-dimensional settings where protected states or access-bearing quantum resources admit richer encodings.

%%%%%%%%%%%%%%%%%%%%%%%%%%%%%%%

\appendices

\section{Minimal Formal Basis of neutralization property}
\label{app:single-use-unsealing}

This appendix recalls the minimal quantum formalism underlying the
custodial argument developed in the paper. 

Following the encrypted-qubit cloning protocol of
Yamaguchi and Kempf~\cite{yamaguchi2026encrypted}, let $A$ denote an
unknown input qubit in state $\rho_A=\ket{\psi}\bra{\psi}$. For $i=1,\ldots,n$, let
$(S_i,K_i)$ be a signal--key qubit pair initially prepared in the Bell state
\[
|\phi\rangle_{S_iK_i}
=
\frac{1}{\sqrt{2}}
\bigl(
|00\rangle_{S_iK_i}
+
|11\rangle_{S_iK_i}
\bigr).
\]
The signal systems $S_i$ become the encrypted clones, while the key
systems $K_i$ \textit{jointly} make up the decryption resource, whereas individually are completely uninformative. The encrypted
cloning operation may be written as a unitary $U^{(n)}_{\mathrm{enc}}$, which acts on \(A\, S_1\cdots S_n\), while leaving the key systems \(K_1,\ldots,K_n\) untouched. In the explicit construction of
Ref.~\cite{yamaguchi2026encrypted}, this unitary 
based on the Pauli operators:
\(
\sigma_0=I,\,
\sigma_1=X,\,
\sigma_2=Y,\,
\sigma_3=Z
\)
is
%%%%
\[
U_{\mathrm{enc}}^{(n)}
=
\frac{1}{2}\sum_{\mu=0}^3 \alpha_\mu^{-1}\sigma_\mu^{(A)}
\otimes\left(\bigotimes_{i=1}^n \sigma_\mu^{(S_i)}\right),
\]
with
$
\alpha_0=1$,
$\alpha_1=\alpha_3=i$, and
$\alpha_2=-i^{\,n+1}$.
However, as suggested in~\cite{gianini2026beyond}, the same structure could be instantiated through different Quantum Secret Sharing architectures; the unauthorized sets need not to be completely blind \cite{gianini2026encrypted,gianini2026full}.

For each signal qubit $S_k$, there exists a decoding operation using
$S_k$ together with the full key register
\[
K = \big(\,K_1,\cdots, K_n\,\big) .
\]
Abstractly, the corresponding unsealing map has the form
\begin{equation}
U_{\mathrm{dec},k}^{(n)}:
S_k K
\longrightarrow
S_k ,
\end{equation}
and 
\begin{comment}
satisfies
\begin{equation*}
\mathcal{D}^{(n)}_k
\left(
\operatorname{Tr}_{\overline{S_kK}}
\left[
U^{(n)}_{\mathrm{enc}}
\left(
\rho_A\otimes
\bigotimes_{i=1}^n
|\phi\rangle\!\langle\phi|_{S_iK_i}
\right)
U^{(n)\dagger}_{\mathrm{enc}}
\right]
\right)
=
\rho_A .
\label{eq:perfect-recovery}
\end{equation*}
%Equation~\eqref{eq:perfect-recovery} 
which 
\end{comment}
expresses perfect recovery from
any one encrypted clone, provided that the full \textit{key register} is
available.

The post-compromise custodial property used in this paper is the
single-use character of the key register. If a trusted clone $S_k$ is
legitimately unsealed using $K$, then the same effective decryption
resource is no longer available for any other encrypted clone. 

After the trusted unsealing of clone \(S_k\), let
\(\rho^{(k)}_{\mathrm{rem}}(\rho_A)\) denote the joint state of all
\textit{remaining}, unopened systems. The single-use character of the decryption
resource is stronger than the mere failure of a second decoder. It means
that the residual state is independent of the original payload:
\begin{equation}
\rho^{(k)}_{\mathrm{rem}}(\rho_A)
=
\tau^{(k)}_{\mathrm{rem}}
\qquad
\text{for all } \rho_A .
\label{eq:residual-no-information}
\end{equation}
%Equivalently,
%\begin{equation}
%\left\|
%\rho^{(k)}_{\mathrm{rem}}(\rho_A)
%-
%\rho^{(k)}_{\mathrm{rem}}(\sigma_A)
%\right\|_1
%=
%0
%\qquad
%\forall \rho_A,\sigma_A .
%\label{eq:trace-distance-no-information}
%\end{equation}
for some fixed state \(\tau_C\) independent of \(\rho_A\). Thus, no subsequent quantum operation acting on
the residual systems can reveal information about the original state.

The corresponding no-information property can be stated in the 
language of complementary quantum channels, or as an
instance of the no-hiding principle~\cite{braunstein2007no}. 

To connect protocol-level description with complementary channel
argument, it is useful to make explicit the induced isometry. Starting
from the input Hilbert space \(\mathcal H_A\), the preparation of the
Bell-pair register embeds the payload into a wider space including the
Bell register
\(
|\Phi\rangle_{SK}
=
\bigotimes_{i=1}^n
|\phi\rangle_{S_iK_i}.
\)
For a chosen trusted clone
\(S_k\), the subsequent unsealing operation selects a \textit{recovered} subsystem
\(B\) (the redeemed clone together with the
decryption resource consumed during unsealing), and a \textit{complementary residual} subsystem \(C\) (the remaining systems, including any stolen dormant clone). At the level of
Hilbert spaces, the corresponding composition can be summarized as
\[
\mathcal H_A
\xrightarrow{\;\rho_A \mapsto \rho_A \otimes
|\Phi\rangle\!\langle\Phi|_{SK}\;}
\mathcal H_A\otimes\mathcal H_{\mathrm{Bell}}
\xrightarrow{\;U_{\mathrm{enc}}^{(n)}\;}
\]\[
\xrightarrow{\;U_{\mathrm{enc}}^{(n)}\;}
\mathcal H_A\otimes\mathcal H_{\mathrm{Bell}}
\xrightarrow{\;U_{\mathrm{dec},k}^{(n)}\;}
\mathcal H_B\otimes\mathcal H_C .
\label{eq:encoding-decoding-isometry-chain}
\]
If there exists a recovery channel \(\mathcal G_B\) such that
\begin{equation}
\mathcal G_B
\left(
\operatorname{Tr}_C
\left[
V \rho_A V^\dagger
\right]
\right)
=
\rho_A
\qquad
\forall \rho_A ,
\label{eq:perfect-recovery-from-B}
\end{equation}
then the complementary subsystem \(C\) is decoupled from the input:
\begin{equation}
\operatorname{Tr}_B
\left[
V \rho_A V^\dagger
\right]
=
\tau_C
\qquad
\forall \rho_A ,
\label{eq:complement-decoupled}
\end{equation}
Thus, after successful legitimate unsealing, the residual systems are
not merely undecodable by the original decoder; they are
completely non-informative about the original payload state. No subsequent  operation acting only on the residual
systems can distinguish which input state was encoded.

The argument in this paper does not depend on the detailed gate-level
implementation of $U^{(n)}_{\mathrm{enc}}$ or of the corresponding
decoder. It depends only on three structural properties of the protocol:
encrypted clones are individually unusable without the key register;
any one clone can be perfectly recovered using the complete noise
register; and the recovery operation consumes the resource required for
further recovery.

\paragraph*{Remarks on quantum storage.}
The abstraction used in this work does not require committing to a
specific physical implementation of quantum share custody. In
an ideal QSS realization, a share may be a quantum subsystem held by a
participant until reconstruction. Physically, this presupposes some form
of quantum storage: a device or medium capable of preserving the relevant
quantum state, or the entanglement needed to reconstruct it, with
sufficient fidelity over the required time scale. Quantum storage is a
central requirement for quantum communication and networked quantum
information processing, but robust long-duration storage remains a
significant technological challenge
\cite{lvovsky2009optical,heshami2016quantum}.

 Depending on the underlying implementation, the distributed unsealing
resource could be realized through persistent quantum shares supported
by quantum-storage devices
\cite{lvovsky2009optical,heshami2016quantum}, short-lived quantum states
used in an online secret-sharing or reconstruction procedure
\cite{hillery1999quantum,lu2016secret}, or hybrid classical--quantum
mechanisms in which classical information reduces, gates, or complements
the quantum resources required for reconstruction
\cite{fortescue2012reducing,lipinska2020verifiable,sun2025classicalquantum}.
The custody models adopted here should be read at the level
of the access structure rather than as a commitment to a specific
storage technology.

%%%%%%%%%%%%%%%%%%%%%%%%%%%%%%%%
\bibliographystyle{IEEEtran}
\bibliography{refs}

\end{document}